\documentclass[aps,twocolumn]{revtex4}

\usepackage{amsmath}
\usepackage{amssymb}
\usepackage{comment}
\usepackage{graphicx}
\usepackage{xcolor}
\usepackage[utf8]{inputenc}
 \usepackage[]{hyperref}

\begin{document}

\title{Technological progress and innovation cascades in Watts–Strogatz networks}

\author{Allan R. Vieira$^1$, 
Carlos Alexandre Gama$^{2,3}$, 
Celia Anteneodo$^1$}  
\affiliation{$^1$Department of Physics, Pontifical Catholic University of Rio de Janeiro, Rua Marquês de São Vicente 225, Gávea, Rio de Janeiro, 22451-900 RJ, Brazil}

\affiliation{$^2$Department of Industrial Engineering, Pontifical Catholic University of Rio de Janeiro, Rua Marquês de São Vicente 225, Gávea, Rio de Janeiro, 22451-900 RJ, Brazil}

\affiliation{$^3$Petrobrás, Rua Henrique Valadares 28, Centro, Rio de Janeiro, 20231-030 RJ, Rio de Janeiro, Brazil}

\begin{abstract}
We investigate how network connectivity and wiring randomness affect technological innovation cascades in a model where agents advance spontaneously or adopt innovations from their neighbors. Extending the model of Guardiola \textit{et al.} to Watts-Strogatz networks, we characterize the maximum progress rate  $\rho_{\max}$ and the corresponding adoption cost $C_{\max}$  as functions of mean degree $k$ and rewiring probability $p$.
We find that the effect of rewiring depends qualitatively on connectivity: increasing $p$ reduces $\rho_{\max}$ in sparsely connected networks, enhances it in highly connected networks, and produces a nonmonotonic response at intermediate connectivity. The maximum attainable progress also varies nonmonotonically across network structures, exhibiting a minimum at intermediate values of $k$ for the parameters investigated. Moreover, the pairs $(C_{\max},\rho_{\max})$ obtained for
different combinations of $k$ and $p$ approximately lie along a common nonmonotonic curve, despite the distinct effects of these two network parameters. These findings demonstrate that long-range links do not necessarily enhance technological progress and that their impact depends on the underlying network connectivity.
 
\end{abstract}

\maketitle

\section{Introduction}
\label{sec:introduction}

Cascades of innovation can arise when the creation of new knowledge or technology by one actor increases the probability of further advances by others. Individual contributions may provide new opportunities, tools, or insights that lower barriers to subsequent innovations, generating chains of technological progress. A familiar example is the sequence of developments from the transistor to integrated circuits, microprocessors, personal computers, and eventually the Internet. More generally, networks of contact, collaboration, and shared resources can facilitate the propagation of knowledge and increase the likelihood of subsequent advances.

An innovation, however, does not necessarily trigger an immediate cascade. Further progress may depend on complementary knowledge, resources, or infrastructure reaching a sufficient level. For example, early developments in neural networks only led to widespread technological applications much later, when advances in computing power and the availability of large datasets removed important constraints. Thus, the propagation of innovation may depend not only on the existence of new ideas but also on the interactions among the agents involved and on the barriers to adopting or building upon previous advances.

Cascading processes have been studied using different approaches. Some models emphasize social mechanisms of influence, such as threshold models of collective behavior~\cite{granovetter} and generalized contagion models~\cite{dodds2004}, whereas others focus on the systemic propagation of technological and organizational changes. An example is the exaptive bootstrapping model~\cite{exaptiveboots}, in which innovations can generate further opportunities through positive feedback mechanisms.

A particularly relevant approach was introduced by Guardiola \textit{et al.}~\cite{PhysRevE.66.026121}, who proposed a model in which each agent, representing a person, firm, or institution, is characterized by a technological level that can be upgraded. Technological advances occur either spontaneously or through interactions with neighboring agents. An agent adopts the technological level of a neighbor when the corresponding technological advantage exceeds a prescribed cost, generating cascades or avalanches of technological progress. The overall progress is characterized by a macroscopic observable that depends on both the technological advance and the size of the resulting avalanches. In this model, the rate of progress exhibits a maximum at an intermediate value of the cost parameter. The model was originally investigated on a one-dimensional ring and on a fully connected network~\cite{PhysRevE.66.026121}.

Subsequent work has considered additional properties of the interacting agents and their connections. Lozano and Arenas~\cite{lozano2007}, for example, investigated the effects of diversity in agent resilience and of homophily in the interaction network. These studies indicate that the structure of interactions can substantially affect the propagation and resilience of innovation cascades. Related work on complex contagion has shown that long-range ties need not always facilitate adoption when reinforcement from multiple contacts is important~\cite{centola2007,centola2010,Kates-Harbeck2023,Eckles2024}. Although the adoption rule considered here is different, these findings provide a broader motivation for examining how the effect of rewiring depends on connectivity. These observations motivate asking how connectivity and the placement of long-range connections jointly affect the rate of technological progress in a model with a different, technology-dependent adoption rule.

In this work, we address this question by extending the innovation-cascade model of Guardiola \textit{et al.}~\cite{PhysRevE.66.026121} to Watts--Strogatz (WS) networks~\cite{watts1998}, allowing the average degree $k$ and the rewiring probability $p$ to be varied systematically. The parameter $k$ sets the mean number of contacts, while $p$ changes their arrangement by replacing local links with longer-range connections. 
We investigate how these two structural parameters affect the dependence of the progress rate $\rho$ on the cost of technological upgrading $C$, with particular emphasis on the maximum progress rate $\rho_{\max}$ and the corresponding cost $C_{\max}$.

Although simplified, the model captures general features of
spreading processes on networks in which propagation depends
on a threshold for adoption. Such processes arise, for example,
in the diffusion of environmentally friendly technologies
among firms, where adoption depends on both technological
barriers and the network of interactions~\cite{ramkumar2022diffusion}.
In our model, spontaneous innovations occur at randomly
selected nodes, without favoring particular agents.
This allows us to focus on how network connectivity and
long-range links influence the propagation of innovation
cascades.

\section{Model}
\label{sec:model}

\subsection{Innovation-cascade dynamics}

In the model introduced by Guardiola \textit{et al.}~\cite{PhysRevE.66.026121}, $N$ agents, representing firms, universities, or other innovation-producing entities, interact through a network of contacts. Each agent $i$ is characterized by a technological level $a_i$, which evolves through two mechanisms: spontaneous technological advances and advances triggered by interactions with neighboring agents.
The dynamics proceeds as follows:

\begin{enumerate}
\item
An agent $i$ is selected uniformly at random and undergoes a spontaneous technological advance,
\begin{equation}
a_i \rightarrow a_i+\Delta_i,
\end{equation}
where the increment $\Delta_i$ is drawn from the exponential distribution
\begin{equation}
p(\Delta)=\frac{1}{\lambda}e^{-\Delta/\lambda}.
\end{equation}
This spontaneous advance initiates an innovation cascade. 

\item
The neighbors of the updated agent are examined. A neighboring agent $j$
adopts the higher technological level, joining the cascade, if
\begin{equation}
a_i-a_j\geq C,
\end{equation}
where $C$ represents the cost, or resistance, associated with technological
adoption. When this condition is satisfied, agent $j$ updates its
technological level according to
\begin{equation}
a_j\rightarrow a_i.
\end{equation}
Each newly updated agent then propagates the innovation to its own neighbors
according to the same rule. The cascade continues until no further adoption
occurs.
\end{enumerate}

Each spontaneous innovation event and its resulting cascade constitute one
elementary update of the dynamics. One Monte Carlo step (MCS) consists of $N$
such events, with each cascade completed before the next spontaneous innovation
is initiated. Thus, one MCS corresponds to $N$ innovation cascades, including
those involving only the initially selected agent.

The avalanche size $s$ is defined as the number of agents whose technological
levels increase during a cascade. The total technological advance associated
with a cascade is defined as
\begin{equation}
A=\sum_{j=1}^{N}\left(a_j^{\mathrm{after}}-a_j^{\mathrm{before}}\right),
\end{equation}
where the superscripts refer to the technological levels immediately before
and after the cascade. The interplay between the adoption cost $C$ and the
network structure determines how far an innovation can propagate.

\subsection{Watts–Strogatz networks}

Whereas the original model considered a one-dimensional ring with $k=2$ as well as a fully connected network~\cite{PhysRevE.66.026121}, here we investigate Watts--Strogatz (WS) networks~\cite{watts1998} with varying mean degree $k$ and rewiring probability $p$. In the standard WS construction, each vertex initially has $k$ neighbors on a regular ring, and each eligible edge is rewired with probability $p$, avoiding self-loops and duplicate edges. Thus, $p=0$ corresponds to a regular ring, and $p=1$ corresponds to a maximally rewired WS network, not necessarily to an Erd\H{o}s--R\'enyi graph. Intermediate values of $p$ can combine local clustering with shorter typical paths~\cite{newmanwatts1999,barratweigt2000}, thus displaying the small-world property.
 
We restrict the analysis to connected interaction networks, so that all agents belong to a single connected component. Network realizations disconnected by rewiring are discarded. The results consequently refer to the WS ensemble conditioned on connectedness, rather than the unrestricted ensemble. This distinction may be especially relevant for sparsely connected networks.
 
Moreover, the standard WS construction requires an even degree $k$. We extended this construction to odd degrees $k=3$ and $k=5$, following an approach similar to that of  Ref.~\cite{BARRIERE2003211}. For $k=3$ ($k=5$), we start from a regular ring in which each node is connected to its one (two) nearest neighbors on either side, yielding an initial degree 2 (4). We then add edges between second- (third-) nearest neighbors according to a periodic pattern, such that each node receives exactly one additional connection. This requires $N$ to be a multiple of 4 (6), respectively. The resulting networks are regular, with degree $k=3$ ($k=5$), and contain only short-range connections.

\subsection{Observables and maximum progress}

Following Ref.~\cite{PhysRevE.66.026121}, we characterize technological progress by

\begin{equation}
\rho = \langle A\rangle/\langle s\rangle,
\label{eq:rho}
 \end{equation}
 where $A$ denotes the  technological advance associated with an avalanche, and $s$ is the number of agents participating in it. 
 The brackets denote averages over the simulated avalanche dynamics and network realizations.
 The ratio of averages in Eq.~\eqref{eq:rho} must be distinguished from the average of the event-wise ratios, $\langle A/s\rangle$. We retain the term ``progress rate'' used in the original model, although $\rho$ is not a rate per unit of physical time.
 
For a fixed network ensemble $(k,p)$,  the cost  $C$ is varied to determine 
\begin{equation}
\rho_{\max}(k,p)=\max_C \rho(C;k,p),
\label{eq:rhomax}
\end{equation}
and the cost at which this maximum occurs,
\begin{equation}
C_{\max}(k,p)=\operatorname*{arg\,max}_C\rho(C;k,p).
\label{eq:cmax}
\end{equation}
Thus, $C_{\max}$ is the cost yielding the largest progress rate for a given network structure; it is not the largest admissible cost.

\subsection{Simulation protocol}

Unless otherwise stated, simulations were performed with $N=1024$ agents. For each choice $(C;k,p)$, 100 independent connected network realizations were generated. For each realization, the dynamics was simulated for 1200 Monte Carlo steps (MCs), discarding the first 1000 MCs for average computations. Observables were averaged  over network realizations and simulation time. 
The system-size dependence was also examined for $N=256$ and $512$.

\subsubsection*{Rescaling of technological increments}

\begin{figure}[t!]
\includegraphics[width=0.99\columnwidth]{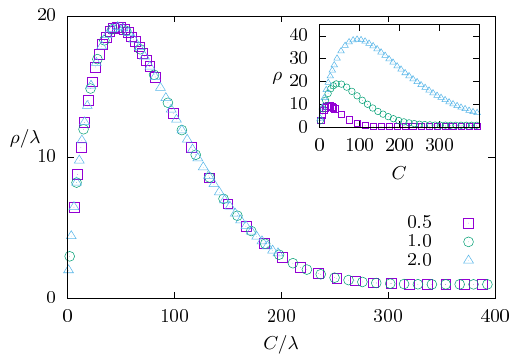}
\caption{Rescaled progress $\rho/\lambda$ versus rescaled cost $C/\lambda$ for a  fully connected network with $N=1024$ and the  indicated values of $\lambda$. 
The inset shows the corresponding unscaled curves  $\rho$ vs $C$.  
}
\label{fig:MF}
\end{figure}

The parameter $\lambda$ sets the scale of the exponentially distributed spontaneous increments. Because the adoption condition compares technological-level differences with the cost $C$, a simultaneous rescaling of all technological levels, increments, and $C$ by the same factor leaves the adoption decisions unchanged, provided no other technological scale enters the dynamics. This motivates the scaling
\begin{equation}
\widetilde{\rho}=\rho/\lambda,
\qquad
\widetilde{C}=C/\lambda.
\label{eq:rescaling}
\end{equation}
The resulting collapse is illustrated in Fig.~\ref{fig:MF} for the fully connected network and is also observed for WS networks. We therefore set $\lambda=0.5$ throughout the subsequent analysis, facilitating comparison with Ref.~\cite{PhysRevE.66.026121}.

\section{Results on WS networks}
\label{sec:results}
 
 We investigate how the progress rate $\rho$ depends on the adoption cost $C$ for different network connectivities and rewiring probabilities. Unless otherwise stated, $N=1024$.

\begin{figure}[b!]
\includegraphics[width=0.99\columnwidth]{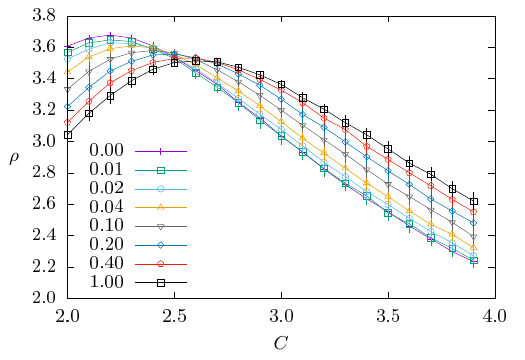}
\caption{Progress rate $\rho$ vs. adoption cost $C$ for connected WS networks with $N=1024$, mean degree $k=2$, and rewiting probabilities $p$  indicated in the legend. 
Symbols show averages and error bars show standard deviations from the simulations.
}
\label{fig:k2}
\end{figure}

Figure~\ref{fig:k2} shows $\rho(C)$ for the sparsely connected case $k=2$. At $p=0$, the regular ring reproduces the qualitative behavior found in Ref.~\cite{PhysRevE.66.026121}. As $p$ increases, the maximum shifts toward larger costs, while its height decreases. Thus, adding long-range links through rewiring does not necessarily improve the maximum progress rate. In this regime, the regular ring yields the largest $\rho_{\max}$ among the values of $p$ investigated.

\begin{figure}[b!]
\includegraphics[width=.99\columnwidth]{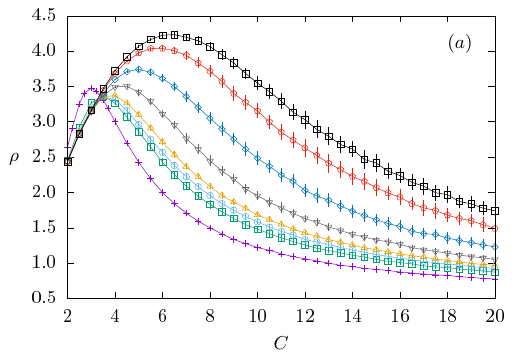}
\includegraphics[width=.99\columnwidth]{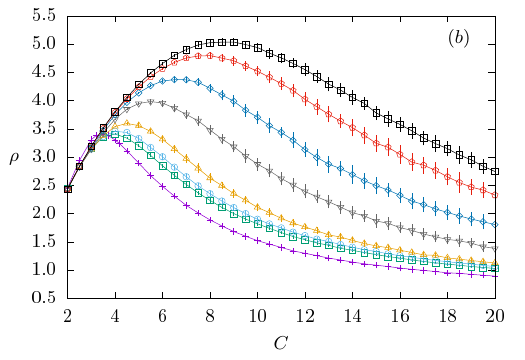}
\caption{Progress rate $\rho$ versus cost $C$ for connected WS networks with $N=1024$ and mean degree (a) $k=4$ and (b) $k=6$. The rewiring probabilities are those considered in Fig.~\ref{fig:k2}.
}
\label{fig:k4e6}
\end{figure}

The effect of rewiring changes as connectivity increases. For $k=4$ [Fig.~\ref{fig:k4e6}(a)], $\rho_{\max}$ varies nonmonotonically with $p$, even though $C_{\max}$ generally shifts toward larger costs. For $k=6$ [Fig.~\ref{fig:k4e6}(b)], increasing $p$ instead enhances $\rho_{\max}$ and shifts its location toward larger $C$. The sign of the rewiring effect therefore depends on the underlying connectivity: randomness is detrimental at low $k$, has a nonmonotonic effect at intermediate $k$, and is beneficial at higher $k$ within the investigated range. 

A possible interpretation involves competing consequences of rewiring. In a sparse ring, replacing a local link with a distant one can interrupt a local propagation pathway, while providing a shortcut to a different region of the network. At larger $k$, redundant local contacts may make such disruptions less consequential, allowing the long-range connections to become more advantageous. This interpretation provides a possible explanation for the observed reversal in the effect of rewiring as connectivity increases.

\begin{figure}[h!]
\includegraphics[width=0.99\columnwidth]{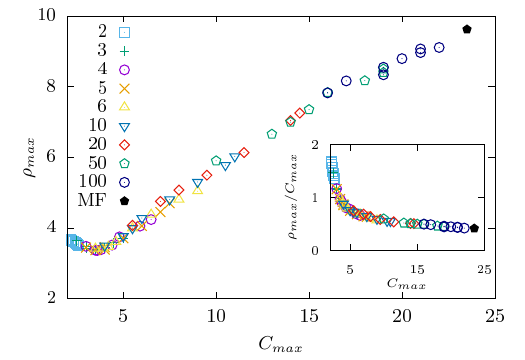}
\caption{
Maximum progress rate $\rho_{\max}$ versus the corresponding cost $C_{\max}$ for connected WS networks with $N=1024$ and the values of $k$ shown in the legend. For each $k$, the points correspond to the eight rewiring probabilities used in Figs.~\ref{fig:k2} and~\ref{fig:k4e6}. The inset displays $\rho_{\max}/C_{\max}$ versus $C_{\max}$.
}
\label{fig:maxima}
\end{figure}

For each network ensemble, maximizing $\rho(C)$ yields the pair $(C_{\max},\rho_{\max})$. Figure~\ref{fig:maxima} displays these pairs for several combinations of $k$ and $p$. Remarkably, the points approximately follow a common, nonmonotonic locus in the $(C_{\max},\rho_{\max})$ plane. Different network structures can therefore yield similar combinations of maximum progress and the cost at which it is attained. This empirical organization does not imply that increasing $k$ is dynamically equivalent to increasing $p$: the dependence of $\rho_{\max}$ on $p$ changes from decreasing to increasing as connectivity increases.

The same representation reveals a minimum of $\rho_{\max}$ at intermediate connectivity, around $k\simeq4$ for the parameters investigated. This is a minimum {across network structures of an already maximized observable}: each value of $\rho_{\max}$ has first been obtained by optimizing the cost $C$ for that network. Hence, intermediate connectivity can be comparatively unfavorable even when the adoption cost is chosen to maximize progress. The nonmonotonic behavior is also visible along the $p=0$ sequence of regular networks, so it is not solely a consequence of conditioning the rewired network ensemble on connectedness.

Since $\rho=\langle A\rangle/\langle s\rangle$,
larger cascades do not necessarily imply greater
technological progress per participating agent.
Thus, network structures that favor widespread
adoption need not be those that maximize $\rho$.

In the standard WS model, clustering generally decreases as rewiring increases. For sufficiently high $k$, however, we observe an increase in $\rho_{\max}$ with $p$, contrary to a simple explanation in which higher clustering invariably favors progress. This suggests that clustering alone is not the dominant determinant of $\rho_{\max}$ in this regime. Because rewiring also changes path lengths and other network properties, the respective structural contributions cannot be disentangled from the present results~\cite{newmanwatts1999,barratweigt2000}. The distinction from reinforcement-based complex-contagion models, where long ties can also have counterintuitive effects~\cite{centola2007,dodds2004}, is that adoption here depends on a technological-level difference rather than repeated social reinforcement.

\begin{figure}[t!]
\includegraphics[width=0.99\columnwidth]{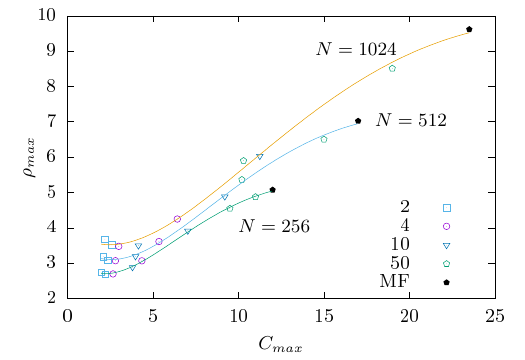}
\caption{Maximum progress rate $\rho_{\max}$ versus the corresponding cost $C_{\max}$ for the system sizes $N=256$, $512$, and $1024$, with network connectivities identified in the legend. Lines guide the eye.}
\label{fig:size}
\end{figure}

Finally, Fig.~\ref{fig:size} illustrates the effect of
system size. The nonmonotonic organization of the maxima
is qualitatively preserved for $N=256$, $512$, and $1024$, although both $\rho_{\max}$ and $C_{\max}$ increase with $N$.
A dependence on system size was also reported by Guardiola
\textit{et al.}~\cite{PhysRevE.66.026121} for both regular
($k=2$, $p=0$) and fully connected networks.
While these quantities scale as $N^{1/2}$ in the mean-field
limit, our results indicate a weaker size dependence
for networks with lower connectivity and rewiring probability.

\section{Final remarks}
\label{sec:final}

We have examined innovation cascades on Watts--Strogatz networks, focusing on the maximum technological progress per participating agent and the adoption cost at which this maximum occurs. The central result is that the influence of rewiring is connectivity-dependent: increasing randomness suppresses maximum progress in sparse networks but enhances it at sufficiently high connectivity, with a nonmonotonic response in between. Thus, long-range links are not intrinsically beneficial for this measure of innovation performance.

The pairs $(C_{\max},\rho_{\max})$ approximately organize along a common, nonmonotonic locus across the network structures considered. In particular, the maximum achievable progress rate displays a minimum at intermediate connectivity. Since each point has already been optimized over the adoption cost, this minimum identifies a comparatively unfavorable class of network structures rather than an unfavorable cost choice within a fixed network. Its qualitative persistence over the finite system sizes investigated further supports the relevance of the effect, although the magnitude of the maxima remains size-dependent.

The contrasting effects of rewiring may reflect a competition between the disruption of local propagation pathways and the benefits of long-range connections, with the balance depending on local connectivity. 

More generally, the extent of innovation diffusion
and the technological progress per participating agent
are distinct quantities. Thus, network structures that favor
larger cascades do not necessarily maximize the progress
rate $\rho$.

Although simplified, the model highlights how the benefits
of increasing connectivity or introducing long-range links
depend on the underlying network structure and on the
quantity used to assess technological progress.

Natural extensions of the present work include the
consideration of heterogeneous or modular networks,
as well as more general innovation-adoption rules.

{\bf Acknowledgments:}
C.A. acknowledges partial financial support from  CNPq (Conselho Nacional de Desenvolvimento Científico e Tecnológico) for Universal 406820/2025-2 and grant 308347/2025-0), and FAPERJ (Fundação de Amparo à Pesquisa do Estado do Rio de Janeiro), for CNE E-26/204.130/2024. 
We also acknowledge partial financial support from CAPES (Fundação Coordenação de Aperfeiçoamento de Pessoal de Nível Superior), through finance code 001. 

{\bf Data availability:}
Data were obtained from numerical simulations and are available under request.

\bibliographystyle{unsrt}
\bibliography{references}

\end{document}